\documentclass[runningheads]{llncs}
\usepackage{graphicx}
\usepackage{amsmath,amssymb} 
\usepackage{color}

\usepackage{subfigure}
\usepackage{amsfonts} 
\usepackage{multirow}
\usepackage{enumitem}

\begin{document}
\title{Eye movement velocity and gaze data generator for evaluation, robustness testing and assess of eye tracking software and visualization tools.} 

\titlerunning{Eye movement velocity and gaze data generator}
%
\author{Wolfgang Fuhl \inst{1} \and
Enkelejda Kasneci\inst{1} }
%
\authorrunning{W. Fuhl and E. Kasneci}
%

\institute{Eberhard Karls University, Sand 14, 72076 Tuebingen, Germany, \email{\{wolfgang.fuhl,enkelejda.kasneci\}@uni-tuebingen.de}}

\maketitle              
\begin{abstract}
Eye movements hold information about human perception, intention, and cognitive state. Various algorithms have been proposed to identify and distinguish eye movements, particularly fixations, saccades, and smooth pursuits. A major drawback of existing algorithms is that they rely on accurate and constant sampling rates, impeding straightforward adaptation to new movements such as microsaccades. We propose a novel eye movement simulator that i) probabilistically simulates saccade movements as gamma distributions considering different peak velocities and ii) models smooth pursuit onsets with the sigmoid function. Additionally, it is capable of producing velocity and two-dimensional gaze sequences for static and dynamic scenes using saliency maps or real fixation targets. Our approach is also capable of simulating any sampling rate, even with fluctuations. The simulation is evaluated against publicly available real data using a squared error. The Matlab code for the simulator can be downloaded at \url{http://ti.uni-tuebingen.de/Projekte.1801.0.html} or used in EyeTrace. 
\end{abstract}

\section{Introduction}
Eye movements hold valuable information about a subject, and his cognitive states~\cite{braunagel2017online,kubler2017subsmatch} and are also important for the diagnosis of defects and diseases of the eyes (many examples can be found in \cite{leigh2015neurology}). Therefore, the detection and differentiation of eye movement types have to be accurate. Most algorithms for eye movement detection apply different dispersion, velocity or acceleration thresholds and validate the detected eye movements based on their duration. This approach seems to be unsatisfactory~\cite{andersson2017one} at its current state. This is partially due to unstable/dynamic sampling rates of eye tracking devices, task-specific sources of noise, the interpolation method applied to the data by the eye tracker, and several more~\cite{cornelissen2002eyelink,duchowski2002breadth}. Depending on the task at hand, different thresholds are proposed in the literature~\cite{holmqvist2011eye}. It is especially difficult to adjust these thresholds for inconsistent sampling rates and noise which is not annotated by the eye tracker. Some commercial eye-tracker differ between tracking the eye and pupil and re-detecting them after a tracking loss, where the latter requires significantly more processing time and thus results in a decreased frame rate. Therefore, the identification of eye movements is still a difficult task; it complicates to confidently generalize research findings across experiments~\cite{andersson2017one}. 

We propose an eye movement simulator to generate data similar to the data of eye-trackers. This is especially useful if algorithms have to be evaluated or data is necessary to test the robustness of software working with eye tracking data. In addition, the generated data can be used to asses visualizations. The proposed simulator currently contains the following features:
\begin{itemize}
	\item Generate velocity profiles of Saccades, Fixations and Smooth Pursuits based on scientific findings.
	\item Generate random sequences following predefined orders.
	\item Generate static and dynamic sampling rates.
	\item Supports any sampling rate.
	\item Generate gaze positions for static images using saliency maps.
	\item Generate new eye tracking data using real data mapped to a saliency map or to real fixation targets.
	\item Generate gaze positions for dynamic scenes like EPIC-Kitchens~\cite{Damen2018EPICKITCHENS} using saliency maps or real fixation targets.
\end{itemize}

\section{Related work}
While there are well-established findings about the gaze signal itself, its synthesis is still challenging. In the Eyecatch~\cite{yeo2012eyecatch} simulator, a Kalman filter is used to produce a gaze signal for saccades and smooth pursuits. While the signal itself was similar to real eye-tracking recordings, the jitter was missing. The first approach for rendering realistic and dynamic eye movements was proposed in \cite{lee2002eyes}, where the main focus was on saccadic eye movements. It also included smooth pursuits, binocular rotations (vergence) and the combination of eye and head rotations. The first data-driven approaches where proposed in \cite{ma2009natural} and \cite{peters2010head}. Both simulate the head and eye movements together in order to generate eye-tracking data. The main disadvantage of \cite{ma2009natural} was that head motion seemed to trigger eye movement. In fact, the head orientation is only changed if the necessary amplitude of the eye is larger than a specific threshold~\cite{murphy2002perceptual} ($\approx 30^\circ$). Another data-driven approach was proposed in \cite{le2012live}, where an automated framework for head motion, gaze, and eyelid simulation was developed. The framework generates data based on speech input using trained Gaussian Mixture Models. While this approach is capable of synthesizing nonlinear data, it only generates unperturbed gaze directions. The approach in \cite{duchowski2015modeling} models eye rotations using specific eye related quaternions for oculomotor rotations as proposed in \cite{tweed1990computing}. The main disadvantage of this approach is that the synthetic eyes cannot be rotated automatically. The approach in \cite{wood2015rendering} produces gaze vectors and eye images to train machine learning approaches for gaze prediction, but does not synthesize realistic eye movements.

All of the aforementioned approaches have their origin in computer graphics with the goal to generate visually realistic head movement and gaze data. The main application of those simulators is to produce realistic interacting virtual humans using parametric models~\cite{andrist2012designing,pejsa2013stylized}. This leads to the disadvantage, that all movements in the generated data are perfect optimal representatives. In reality, the raw gaze data in eye movements contains noise introduced either through actual movements such as microsaccades or inaccuracies of the used eye-tracker. The first approach to simulate a realistic scan path, i.e., a sequence of fixations and saccades, on static images was proposed in \cite{campbell2014saliency}. They use a saliency map together with a unified Bayesian model to generate realistic random walks over a stimulus. A pure gaze data simulation approach including noise was proposed in \cite{duchowski2015eye}. Based on this approach, \cite{duchowski2016eye} further improves the noise synthesis by simulating jitter as a normal distribution.

Our approach combines the before mentioned publications by simulating velocity profiles of three eye movement types and map them to saliency maps. In contrast, our approach is capable of simulating noise as uniform and normal distribution and also allows to produce any sampling rate (static and dynamic). We also extended the mapping functionality to allow the usage of real fixation targets and use dynamic scenes instead of only static images.

\section{Simulation}
\begin{figure}[h]
	\centering\includegraphics[width=0.8\textwidth]{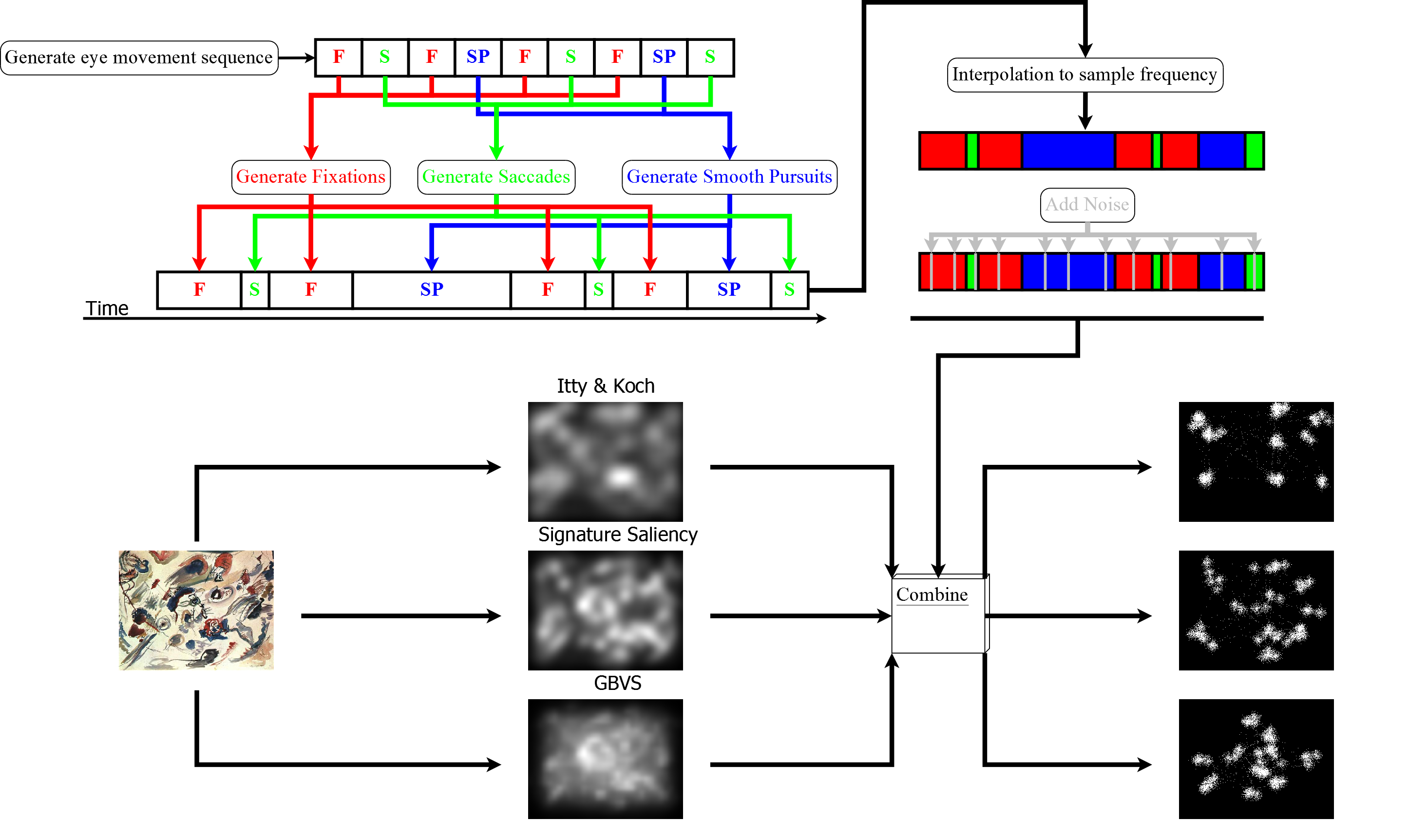}
	\caption{Work-flow of generating eye movement data. First, a sequence of eye movement types is generated. In the second step, a model of each eye movement type is generated (F: Fixation, S: Saccade, SM: Smooth pursuit). This model allows for an almost infinite sampling rate, which is in the next stage interpolated to a target sampling rate (Red: Fixation, Green: Saccade, Blue: Smooth pursuit). Finally, noise is added on top of the signal (gray). The last stage is the mapping of the generated sequence to target locations.}
	\label{fig:workflow}
\end{figure}
The entire work-flow of the simulator is shown in Figure~\ref{fig:workflow}. Generating an eye movement velocity profile is done in four steps. The first step chooses a sequence of eye movement types (Fixation, Saccade, Smooth pursuit) without any time or velocity constraints. Afterward, each movement type in this sequence is assigned a velocity profile generated by preliminary set parameters. The mathematical model behind these profiles allows sampling at an extremely high, almost arbitrary rate. The target sampling rate is obtained by interpolating the computed frequency, which also allows for dynamically adjusting the target sampling rate. In the next step, noise is added which represents measurement errors or blinks. The generated velocity profile is then mapped to two-dimensional locations on a stimulus which are created using saliency maps (\cite{harel2007graph,itti1998model,hou2012image}) or real fixation targets. This also allows to remap real eye tracking data or use a dynamic stimulus taking into account the duration of the individual eye movement types. Each step of this eye movement simulator is described in the following subsections in more detail.
The simulator also includes a random walker generator to model fixation direction~\cite{engbert2011integrated}; saccade and smooth pursuit directions are generated randomly (but consistently within a movement) since this is stimuli- and task-dependent.

\subsection{Eye movement sequence}
Generating a sequence of eye movement types can be done either by sampling from a uniform distribution, setting it manually, or by following construction constraints. In case of the uniform distributed eye movements, the generator script randomly selects between three types of eye movements. If the amount of each type is specified a priori, the probability is automatically adjusted. This means that after each insertion the probabilities are computed based on the remaining quantity of each type to favor higher quantities. This process can also be constrained, e.g., by forcing the algorithm to insert a saccade after each fixation or before a smooth pursuit.

\subsection{Fixation}
Fixations are generated based on two probability distributions which can be specified and parametrized. The first distribution determines its duration, the second the consistency of the fixation. For the duration and consistency, the minimum and maximum can be set. As distributions, the simulator provides Normal and Uniform random number generation. For the Normal distribution, the standard deviation can be specified. consistency describes the fluctuations in the velocity profile and is used as such in the entire document.
\begin{figure}[h]
	\centering
	\subfigure[]{\includegraphics[width=0.48\textwidth]{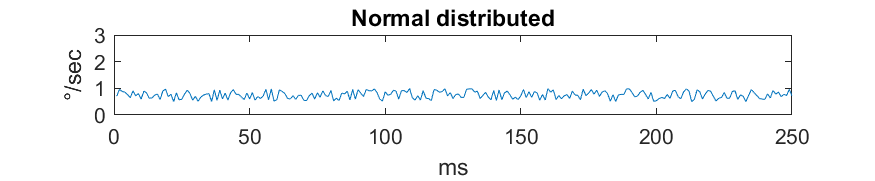}}
	\subfigure[]{\includegraphics[width=0.48\textwidth]{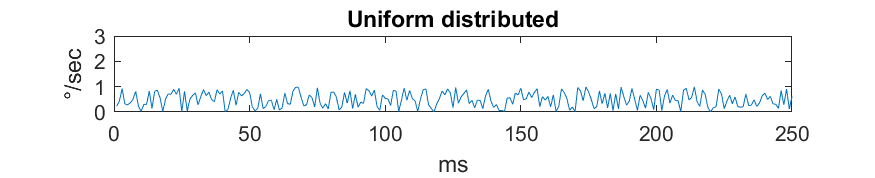}}
	\caption{Generated fixation based on a Normal (a) and Uniform (b) distribution.}
	\label{fig:fixexample}
\end{figure}

In Figure~\ref{fig:fixexample}, two artificially generated fixations are shown. The consistency was set to one degree per second and the standard deviation for the normal distribution to two
(Figure~\ref{fig:fixexample} (a)). As can be seen in the figure, the Uniform distribution looks more similar compared to real data although we have set the consistency very high with one degree per second.

\subsection{Saccade}
The most complex part of the eye movement generator is the saccades. For the length, we follow the same approach as for the fixations, in which a minimum and maximum length have to be set. The selectable distributions are Normal and Uniform. The result of the length also influences the maximum speed of the saccade. Therefore, the two random numbers are multiplied (both in the range between zero and one).
This means that shorter saccades are limited to lower maximal velocities. To generate the velocity profile, minimum, maximum and the distribution type have to be set.

The most characteristic property of a saccade is its velocity profile. In our simulator, this is generated as a Gamma distribution. Therefore, the minimum and maximum skewness have to be specified. In \cite{van1987skewness} it was found that the Gamma function can be considered suitable to approximate saccade profiles (yet not perfect). To achieve more realistic data, a consistency minimum, maximum and distribution can be specified. This generates the jitter along the velocity profile.

\begin{figure}[h]
	\centering
	\subfigure[]{\includegraphics[width=0.48\textwidth]{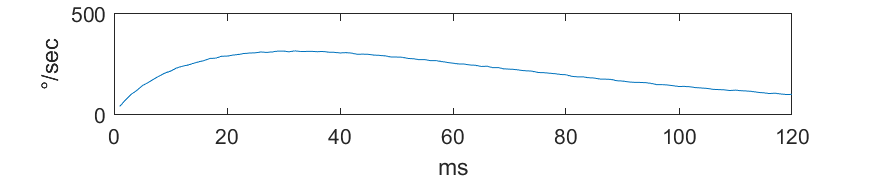}}
	\subfigure[]{\includegraphics[width=0.48\textwidth]{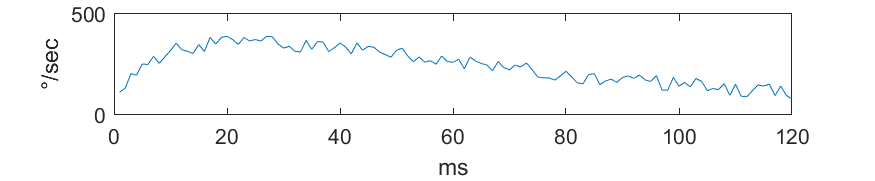}}
	
	\subfigure[]{\includegraphics[width=0.48\textwidth]{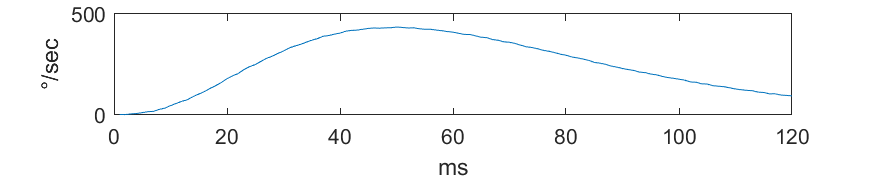}}
	\subfigure[]{\includegraphics[width=0.48\textwidth]{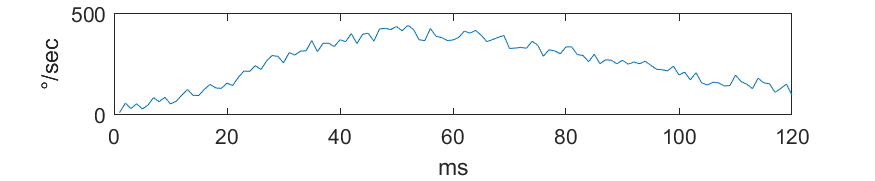}}
	\caption{Generated saccades with jitter (b,d) and without (a,c). For (a) and (b), the distribution was skewed to the left. In (c) and (d), Gamma distribution was only slightly skewed.}
	\label{fig:sacexample}
\end{figure}
Figure~\ref{fig:sacexample} shows some generated saccades of fixed length. We simulated two large and two slightly left skewed saccades. The maximum velocity was selected from a range between $300$ and $500$ degrees per second.
As can be seen from the Figure, the profile contains on- an offset of a saccade. The profile itself is smooth and follows the Gamma distribution. Post-saccadic movement is as of now missing in the simulator. In Figure~\ref{fig:sacexample}(b) and (d), a small amount of jitter was added to simulate measurement inaccuracy. This usually occurs through the approximation on image pixels or ellipse fit inaccuracy in pupil detection.

\subsection{Smooth pursuit}
For generating smooth pursuits we also simulate the onset following the findings in \cite{ogawa1998velocity}. The authors did not provide a final function for the description of the velocity profile but visualized and described it precisely. The shape of the onset of a smooth pursuit follows a nonlinear growing function similar to the sigmoid function. While this equation is not scientifically proven, our framework allows to simply replace it once a better model is available. The most complex part of the pursuit model is the onset, followed by a regular movement.

The parameters that can be specified are the minimum and maximum length together with their distribution type. For the velocity and the length of the onset, the same parameters can be adjusted. To include the measuring error, the consistency parameters are also configurable. For the pursuit itself, we included linear growing, decreasing and constant profiles. In case of the growing, again the minimum, maximum and consistency function can be specified.
\begin{figure}[h]
	\centering
	\subfigure[]{\includegraphics[width=0.16\textwidth]{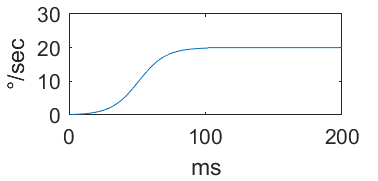}}
	\subfigure[]{\includegraphics[width=0.16\textwidth]{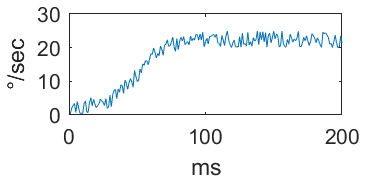}}
	\subfigure[]{\includegraphics[width=0.16\textwidth]{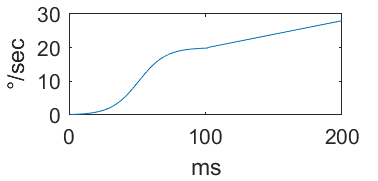}}
	\subfigure[]{\includegraphics[width=0.16\textwidth]{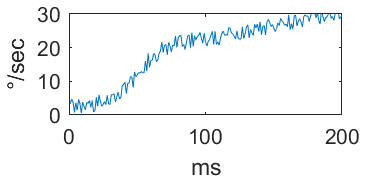}}
	\subfigure[]{\includegraphics[width=0.16\textwidth]{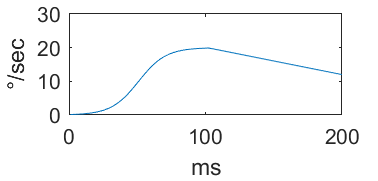}}
	\subfigure[]{\includegraphics[width=0.16\textwidth]{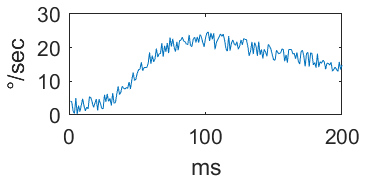}}
	\caption{Generated smooth pursuits with jitter (b,d,f) and without (a,c,e). For (a) and (b), the pursuit movement was constant. In (c,d) and (e,f) it was linear increasing and decreasing.}
	\label{fig:smexample}
\end{figure}
Figure~\ref{fig:smexample} shows simulated smooth pursuits. For the visualization of the linear decreasing and increasing function, extreme values were used. The first column shows a smooth pursuit for a constantly moving object, which is often observed in laboratory experiments. The increasing and decreasing profiles are for objects which move further away or come closer to the subject with a constant speed. Other profiles may occur in real settings too, where the object has a slightly varying speed but these are future extensions of the generator and not part of this paper.

\subsection{Sampling}
\label{sec:dwsampling}
After generating and linking the eye movements, they have to be interpolated to a sampling rate. This is necessary to simulate different recording frequencies. Here it is important to mention that not all modern eye trackers record at a constant frequency. On the one hand, image acquisition rates can vary depending on illumination changes that affect the aperture time of the camera and timestamps generated by the eye-tracker can vary in accuracy. On the other hand, image processing time, e.g. for eye and pupil detection, are not necessarily constant and might change depending on how easy the pupil can be identified. For example, detection of the pupil is usually more time-consuming than keeping track of a previously detected pupil. Some systems, especially when running on mobile devices, may run into a state where frames are dropped in order to maintain real-time performance. We found systems where the timestamps are generated by the CPU time (which may be inaccurate for fast sampling rates) and even timestamps that are generated after image processing. Therefore, our simulator is capable of simulating varying sampling rates. The parameters for this step are the minimum and maximum sampling rate and also the consistency function. The interpolation itself computes the mean of all values from the last sampling position to the new sampling position.

\begin{figure}[h]
	\centering
	\subfigure[]{\includegraphics[width=0.48\textwidth]{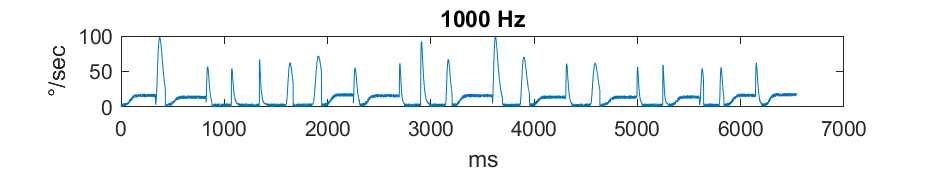}}
	\subfigure[]{\includegraphics[width=0.48\textwidth]{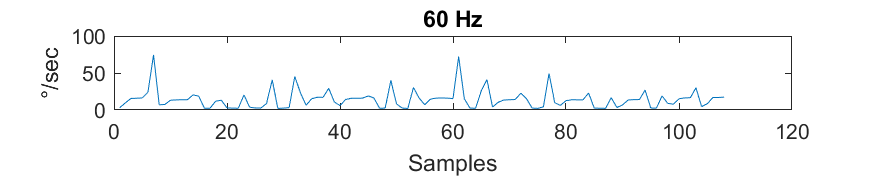}}
	
	\subfigure[]{\includegraphics[width=0.48\textwidth]{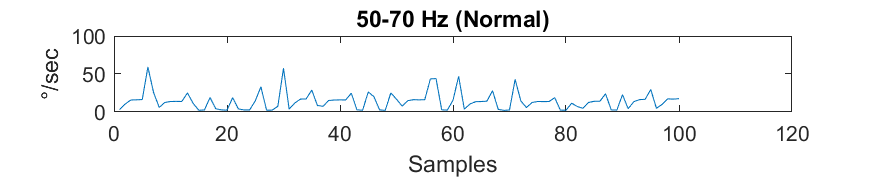}}
	\subfigure[]{\includegraphics[width=0.48\textwidth]{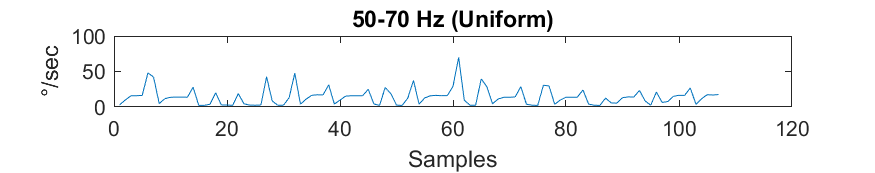}}
	\caption{Generated velocity profile of an eye movement sequence (a). In (b), the data is sampled at 60Hz without variations. (b) and (c) vary between 50 and 70 Hz with the Normal and the Uniform distribution.}
	\label{fig:pathexample}
\end{figure}
In Figure~\ref{fig:pathexample}(a) a generated velocity profile is shown. The initial sampling frequency was set to 1000 Hz but any other sampling rate is possible. For (b), a constant sampling frequency of 60 Hz was used. In (c), the sampling frequency varies between 50 and 70 Hz (with a mean of 60 Hz), wherein the Normal distribution was used as the random number generator. It differs significantly from the constant sampling rate in (a) and also has a different length. For (d), the sampling frequency also varied between 50 and 70 Hz with the difference that the Uniform distribution was used as the random number generator. The length is therefore similar to the constant sampling rate but it still differs especially for the saccadic peeks.

\subsection{Noise}
For generating noise, two distributions are used: one for the location where to place the noise in the data and the second for the velocity change to apply. Therefore, the user has to specify the types for both distributions and the minimum and maximum velocity of noise. The amount of noise is specified as a percentage of the samples that should be influenced.
\begin{figure}[h]
	\centering
	\subfigure[]{\includegraphics[width=0.32\textwidth]{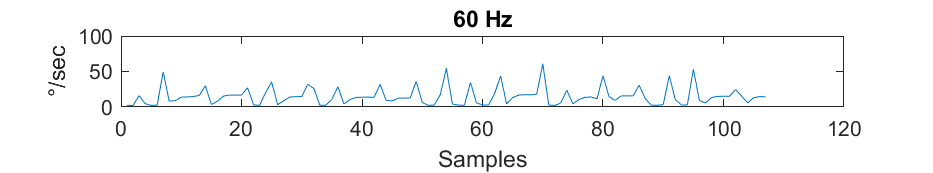}}
	\subfigure[]{\includegraphics[width=0.32\textwidth]{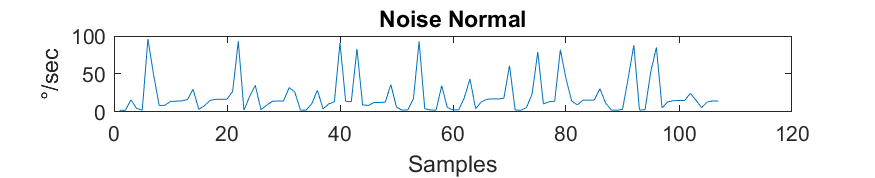}}
	\subfigure[]{\includegraphics[width=0.32\textwidth]{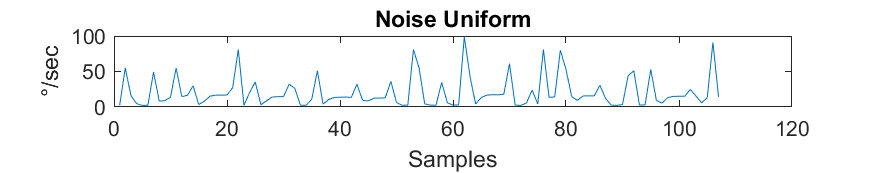}}
	\caption{Generated velocity profile of an eye movement sequence (a). In (b), noise is added based on a Normal distribution and in (c) a Uniform distribution was used.}
	\label{fig:noiseexample}
\end{figure}

Figure~\ref{fig:noiseexample} shows two types of Noise added to the velocity profile shown in (a). The amount of noise added was 10\%. For the Normally distributed noise in (b) it can be seen that the peaks are mostly high. In comparison to it, the Uniform distributed noise in (c) produces more peaks of different heights.

\subsection{Mapping}
\begin{figure}[h]
	\centering\includegraphics[width=0.8\textwidth]{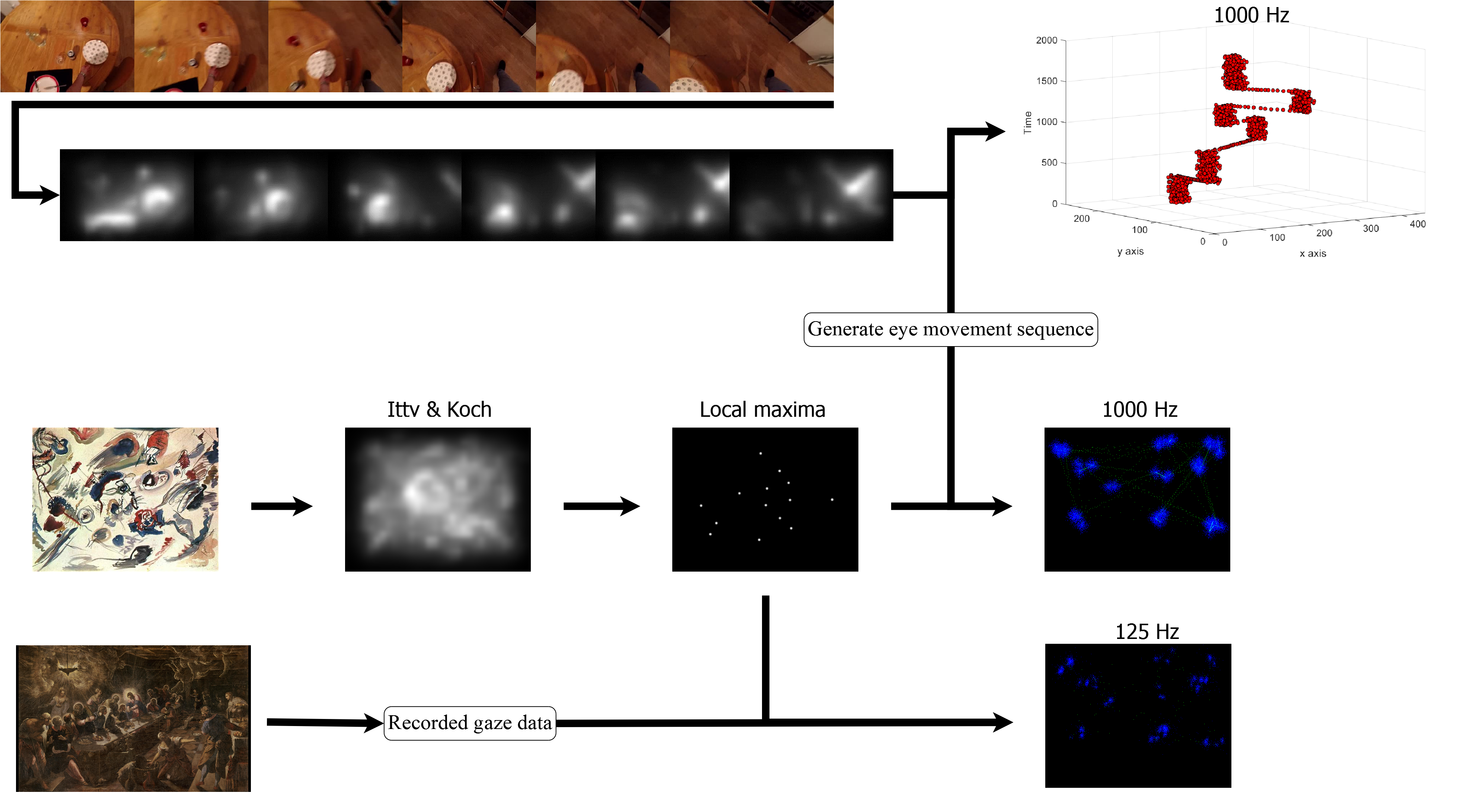}
	\caption{Scenarios of mapping eye movement data. First, a video sequence is converted to saliency maps and afterward gaze data is mapped to them based on the frame rate. The central part shows the mapping of generated or real data to a new stimulus image. At the bottom real data is mapped to the same stimulus based on fixation locations and saccade durations. The upper series of images is taken from EPIC kitchen~\cite{Damen2018EPICKITCHENS}.}
	\label{fig:mappingflow}
\end{figure}
The mapping function of the simulator is used to produce spatial data out of the velocity profiles (Figure~\ref{fig:mappingflow}). This functionality can be used for simulated and real data (fixations, saccades, smooth pursuits). Therefore, possible fixation targets have to be identified for which our simulator includes three saliency maps (\cite{harel2007graph,itti1998model,hou2012image}). As locations, the local maxima of these saliency maps are used. In addition, a small random shift of the local maxima is also included as the possible target to simulated close consecutive fixations. Afterward, a type of eye movement is selected and randomly generated by a predefined parameter set including a duration range. For saccades and smooth pursuits, this duration is interpolated to the distance between the last and new fixation target taking into account the speed of the individual sample points. In addition, a maximal deviation range can be defined based on which the gaze points differ to the straight line between both positions. For fixations, the scattering is generated based on the deviation parameter (inaccuracy of the measurement simulation). For smooth pursuits, both approaches are used to map the velocity profile between to locations.
Since it is possible to generate a velocity profile out of real data to the aforementioned approaches can also be used to map real fixations, saccades, and smooth pursuits to new stimulus images. For the generation of new data out of real data for the same stimulus image, we propose to randomly select an eye movement type out of the real data and use the centers of all fixations as possible targets. An example can be seen at the bottom of Figure~\ref{fig:mappingflow}.

For dynamic scenes, the generated data can be mapped based on the same approach (local maxima of saliency maps). The only thing that differs is that the local maxima are time-dependent. This means that for a saccade the two locations have to be selected out of two different sets of local maxima which are computed based on the timestamps of frames in the video (Figure~\ref{fig:mappingflow} top).

\section{Evaluation}
\begin{figure}[ht]
	\centering
	\subfigure[]{\includegraphics[width=0.3\textwidth]{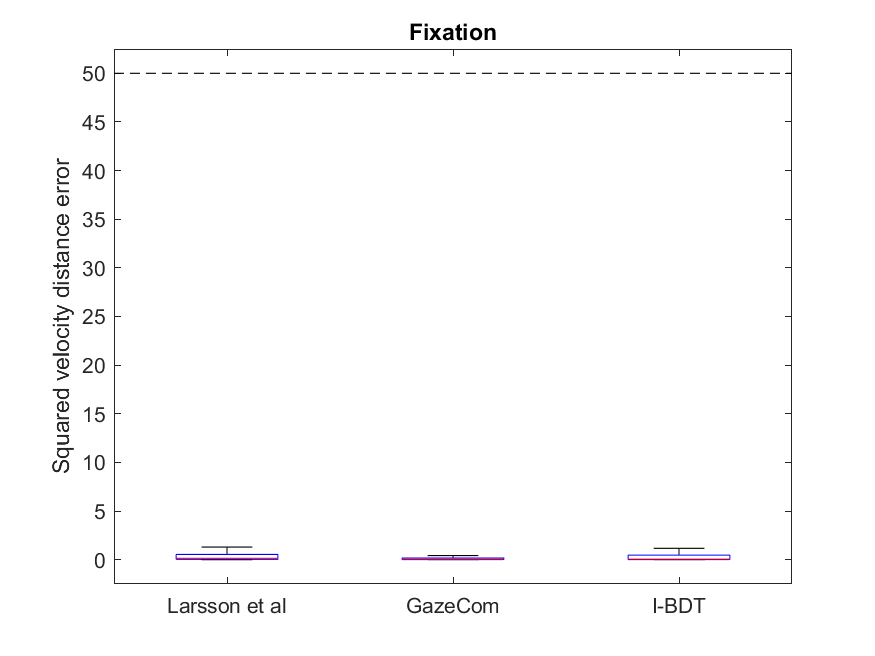}}
	\subfigure[]{\includegraphics[width=0.3\textwidth]{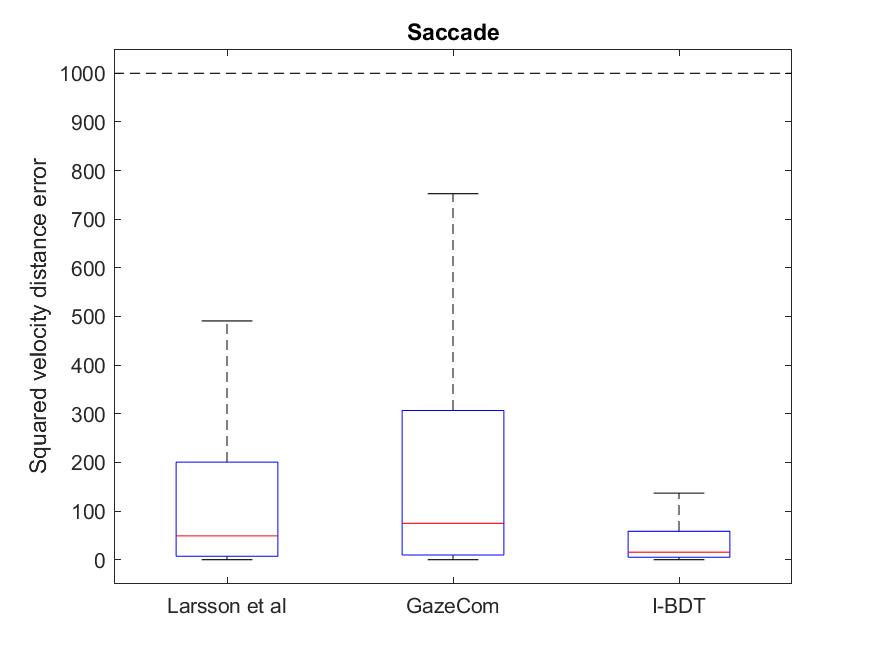}}
	\subfigure[]{\includegraphics[width=0.3\textwidth]{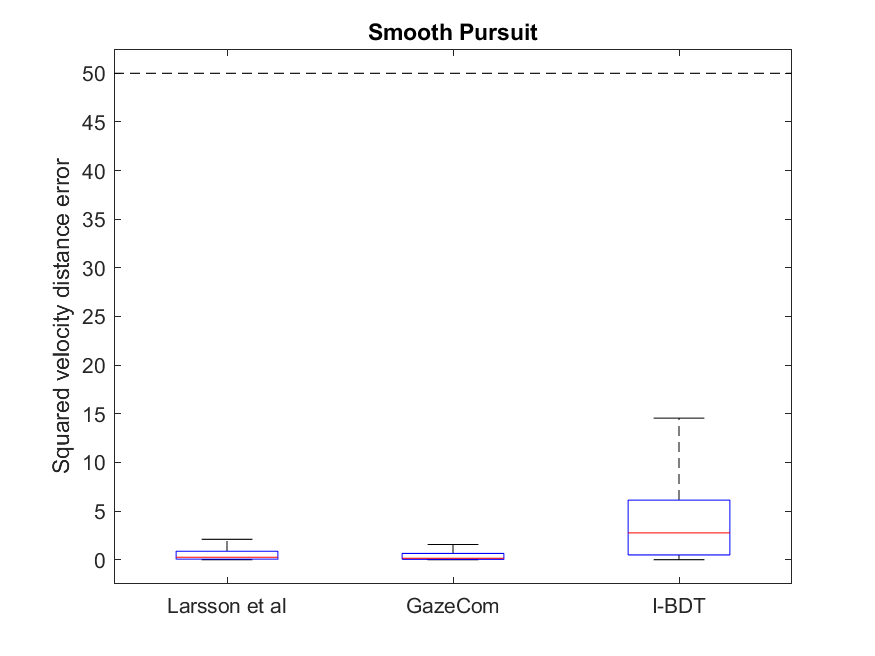}}
	\caption{Squared velocity error for the simulation per data set.}
	\label{fig:eval_simu}
\end{figure}
Figure~\ref{fig:eval_simu} shows the per sample point squared error as whisker plots of our simulator in comparison to the publicly available datasets~\cite{larsson2013detection,spdetectionsite,StAgDo17,santini2016bayesian}. The error was computed based on the squared difference between each sample. Therefore, we simulated each fixation, saccade, and smooth pursuit ten times with the same length as in the available data sets. For a fixation, the simulator got the information of the mean velocity and the standard deviation to generate a profile. The information of a saccade was the peak velocity and the position of this peak. For smooth pursuits, the simulator got the information of the mean velocity and the standard deviation. 

As can be seen in Figure~\ref{fig:eval_simu}(b), the error for saccades was the largest. This is due to the noisy signal which is due to the inter sample velocity computation.
\begin{figure}[htb]
	\centering
	\subfigure[]{\includegraphics[width=0.3\textwidth]{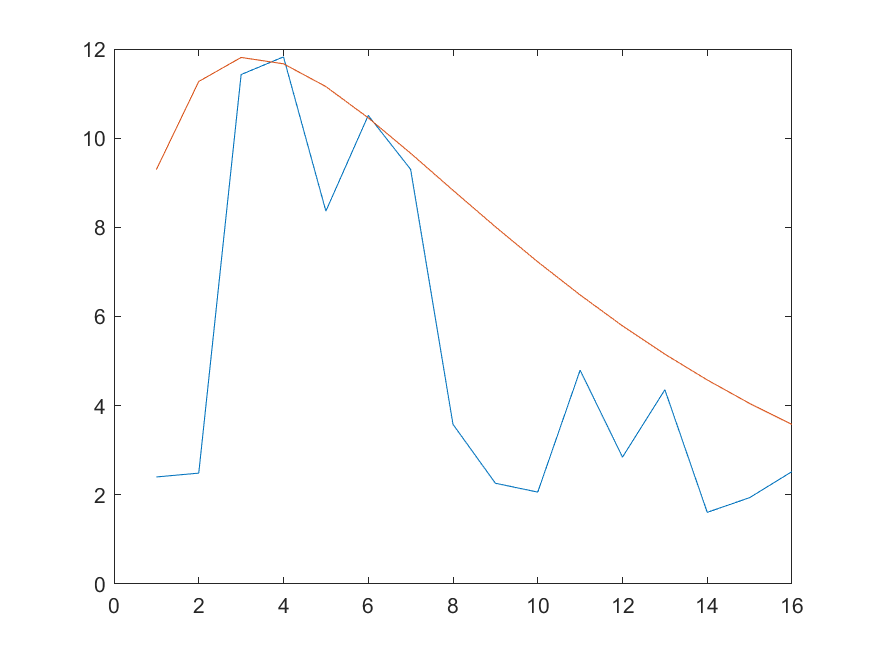}}
	\subfigure[]{\includegraphics[width=0.3\textwidth]{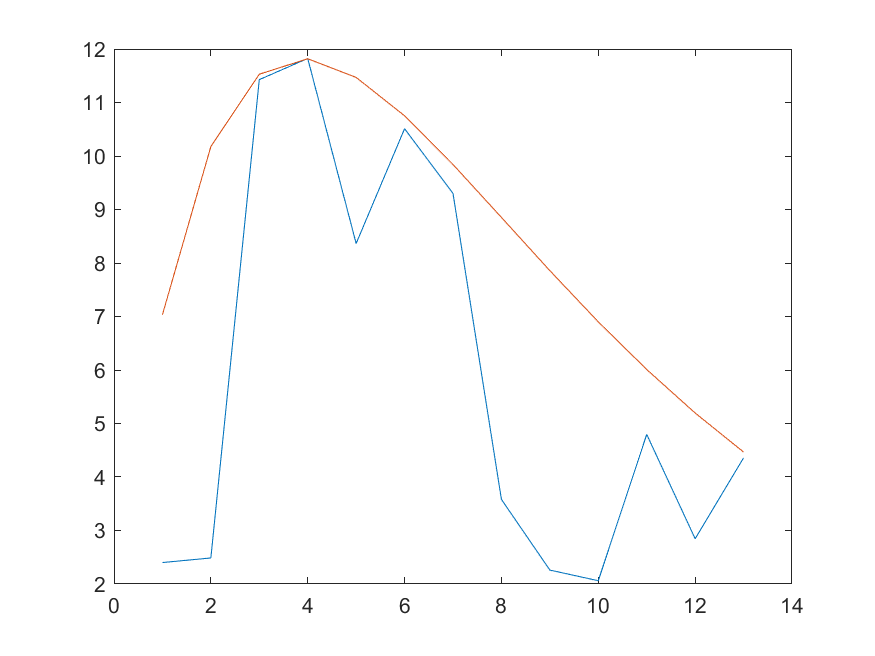}}
	\subfigure[]{\includegraphics[width=0.3\textwidth]{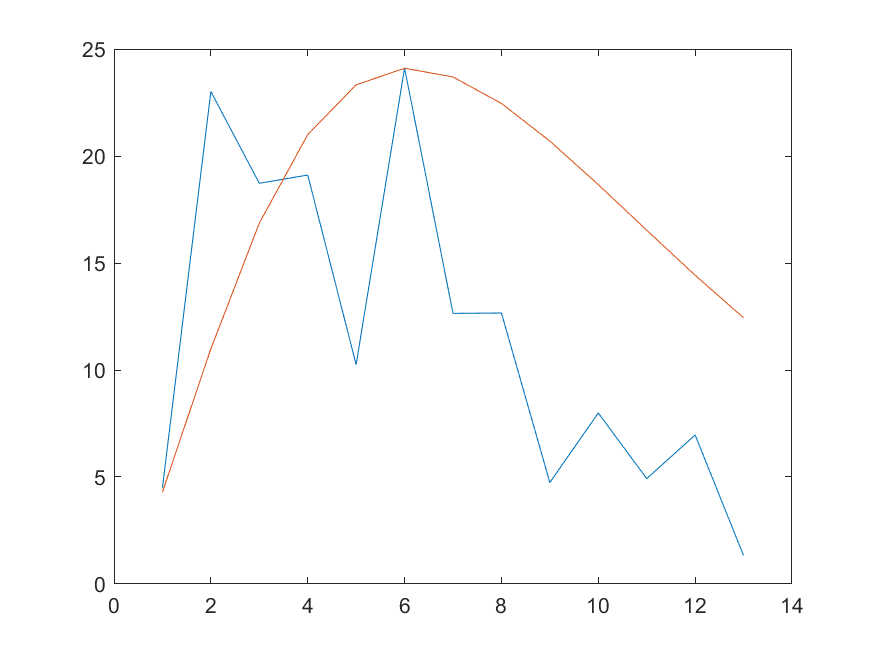}}
	\caption{Saccades with a high squared error. Red is the simulation and blue is the real data.}
	\label{fig:eval_sacc}
\end{figure}
Figure~\ref{fig:eval_sacc} shows some saccades which produced high squared errors. The red line corresponds to the simulation result, whereas the blue line corresponds to the real data. As can be seen, the course of the velocity profile is well simulated, which is well in line with previous findings in~\cite{van1987skewness}. The high errors originate mainly from measurement inaccuracies in the real data. This also highlights the difficulty in detecting eye movements in such a signal. For the data set from \cite{santini2016bayesian}(I-BDT), the error for saccades was lowest. This is due to the low sampling rate of the used eye tracker (30 Hz), for which large fluctuations do not occur. This is similar to smoothing or using multiple samples for the velocity computation. In contrast, the smooth pursuits error was the largest in the I-BDT data set. This is because in such low sampling rates the onset of a smooth pursuit is hardly represented. Our simulator is capable of simulating this (sampling~\ref{sec:dwsampling}) but for the evaluation, it was not used. We only used the generators to simulate the eye movements.

\section{Conclusion}
We proposed a novel eye movement simulator which is capable of creating mappings to static images and dynamic scenes based on saliency maps or real fixation targets. Optionally the framework is also capable of remapping real eye tracking data onto new stimuli or generate a new scan path based on real data for the same stimuli. In addition, it can generate data for any static and dynamic sampling rate. The currently included eye movement types are Fixations, Saccades and Smooth Pursuits which can be parameterized. Variations and noise can be generated using different distributions for noise, sampling shift, eye tracker accuracy etc. Further research will be the extension of the simulator to be also capable of generating post saccadic, optokinetic and vestibular-ocular movement. In addition, smooth pursuits have to follow an object for which we want to include a key point registration and detection to compute possible locations for this type of eye movements in videos.

%
%
%
%
%
%
 \bibliographystyle{splncs04}
 \bibliography{egbib}
\end{document}